\documentclass[pre,groupaddress,twocolumn,nofootinbib,floatfix,11pt]{revtex4}

\usepackage{amsmath,amsfonts,amssymb}
\usepackage{graphicx}
\usepackage{hyperref}
\usepackage{verbatim}
\usepackage{comment}
\usepackage{setspace}

\usepackage[letterpaper, foot=10cm, left=0.75in, right=0.75in, top=0.75in, bottom=0.75in]{geometry}
\begin{document}
\singlespacing

\title{\Large Exploiting imperfections: Directed assembly of surface colloids via bulk topological defects}





\author{Marcello Cavallaro Jr.}
\thanks{These authors contributed equally to this work.}
\affiliation{The Laboratory for Research on the Structure of Matter, University of Pennsylvania, 3231 Walnut Street, Philadelphia, Pennsylvania 19104, USA}
\author{Mohamed A. Gharbi}
\thanks{These authors contributed equally to this work.}
\affiliation{The Laboratory for Research on the Structure of Matter, University of Pennsylvania, 3231 Walnut Street, Philadelphia, Pennsylvania 19104, USA}
\author{Daniel A. Beller}
\thanks{These authors contributed equally to this work.}
\affiliation{The Laboratory for Research on the Structure of Matter, University of Pennsylvania, 3231 Walnut Street, Philadelphia, Pennsylvania 19104, USA}
\author{Simon \v{C}opar}
\affiliation{The Laboratory for Research on the Structure of Matter, University of Pennsylvania, 3231 Walnut Street, Philadelphia, Pennsylvania 19104, USA}
\author{Zheng Shi}
\affiliation{The Laboratory for Research on the Structure of Matter, University of Pennsylvania, 3231 Walnut Street, Philadelphia, Pennsylvania 19104, USA}
\author{Tobias Baumgart}
\affiliation{The Laboratory for Research on the Structure of Matter, University of Pennsylvania, 3231 Walnut Street, Philadelphia, Pennsylvania 19104, USA}
\author{Shu Yang}
\affiliation{The Laboratory for Research on the Structure of Matter, University of Pennsylvania, 3231 Walnut Street, Philadelphia, Pennsylvania 19104, USA}
\author{Randall D. Kamien}
\affiliation{The Laboratory for Research on the Structure of Matter, University of Pennsylvania, 3231 Walnut Street, Philadelphia, Pennsylvania 19104, USA}
\author{Kathleen J. Stebe}
\thanks{To whom correspondence should be addressed. E-mail: kstebe@seas.upenn.edu}
\affiliation{The Laboratory for Research on the Structure of Matter, University of Pennsylvania, 3231 Walnut Street, Philadelphia, Pennsylvania 19104, USA}

\begin{abstract} 
We exploit the long-ranged elastic fields inherent to confined nematic liquid crystals to assemble colloidal particles trapped at the liquid crystal interface into reconfigurable structures with complex symmetries and packings.  Spherical colloids with homeotropic anchoring trapped at the interface between air and the nematic liquid crystal 5CB  create quadrupolar distortions in the director field causing particles to repel and consequently form close-packed assemblies with a triangular habit. Here we report on complex, open structures organized via interactions with defects in the bulk.  Specifically, by confining the nematic liquid crystal in an array of microposts with homeotropic anchoring conditions, we cause defect rings to form at well-defined locations in the bulk of the sample.  These defects source elastic deformations that direct the assembly of the interfacially-trapped colloids into ring-like assemblies, which recapitulate the defect geometry even when the microposts are completely immersed in the nematic.   When the surface density of the colloids is high, they form a ring near the defect and a hexagonal lattice far from it.  Since topographically complex substrates are easily fabricated  and liquid crystal defects are readily reconfigured, this work lays the foundation for a new, robust mechanism to dynamically direct assembly over large areas by controlling surface anchoring and associated bulk defect structure. 

\end{abstract}






\maketitle
{\footnotesize  \noindent topology $\vert$ 5CB $\vert$ elastic interaction $\vert$ nematic $\vert$ micropost  }\\
\\
{\footnotesize \noindent Abbreviations: LC, liquid crystal; 5CB, 4-cyano-4'-pentylbiphenyl; DMOAP, 3-(Trimethoxysilyl)propyl-octadecyldimethylammonium chloride; POM, polarized optical microscopy; FCPM, fluorescence confocal polarization microscopy; LdG, Landau-de Gennes}

\section*{Significance Statement}

\noindent In this research, we develop new means of directing colloids at an interface to assemble into complex configurations by exploiting defects in a liquid crystal (LC). Through confinement of a nematic LC over a topographically patterned surface, we demonstrate the formation of defects at precise locations in the LC bulk. These defects source elastic distortion fields that guide the assembly of colloids constrained to the LC-air interface. This work significantly extends prior work in which LCs confined in film or droplet geometries guide colloidal assembly beyond simple triangular lattices and chains.  Here, we demonstrate colloidal assembly at precise locations, with particle rich and poor regions, determined remotely by defects deliberately seeded in the LC bulk. Experimental results are supported by numerical and analytic investigation.

\section*{Introduction}

\noindent Classically, the bulk of a material system is where the action is: the interface is oft relegated to a set of ``boundary conditions." However, crystal faceting \cite{wulff}, the quantum hall effect \cite{TKNN}, and even the AdS-CFT correspondence \cite{magoo} fundamentally reverse this relationship -- the bulk properties can be read off from their effects on the boundaries. In this contribution, we demonstrate migration and organization of colloids constrained to a liquid crystal-air interface, driven remotely by the elastic distortion created by the presence of
topological defects in the liquid crystalline bulk. Just as phantoms are employed in magnetic resonance imaging \cite{resonance90}, it is necessary for us to prepare bulk defects in known configurations in order to verify our bulk/boundary connection. To do this, we prepare a substrate patterned with microposts that, with appropriate surface treatment, seed a reproducible defect complexion. Colloidal spheres on the interface experience an attraction to the regions above the submerged defects, as well as an elastic repulsion from each other, leading to complex new assemblies. The long range of these elastic interactions  allows defects in the bulk nematic phase far below the interface to direct assembly at the interface. Other recent work on producing ordered arrangements of particles at liquid crystal interfaces beyond simple triangular lattices -- such as chains \cite{GharbiSoftMatter}, stripes \cite{moreno2012liquid}, and dense quasihexagonal lattices \cite{nych2007coexistence} -- have focused on confining the nematic in thin film or droplet geometries and varying the surface coverage fraction.  Our sensitive control over substrate topography provides the ability to tune the defects' positions and their influence on the interface, offering a route to tunable non-trivial colloidal assemblies.

To understand the bulk's influence on the boundary, it is essential for us to first know what is on the inside.  To that end, we have carefully characterized, simulated, and analyzed the response of the nematic liquid crystal, 5CB, to the micropost-patterned substrate.  We begin with a description of the sample preparation, our microscopy results, and an analytic approximation to characterize the defect complexion.  We corroborate this with numerical modeling and find, to our surprise, that the connection between the boundary conditions and the substrate topography is not always entirely geometric.  Finally, we present our main finding: the migration and arrangement of colloids at the surface in response to disclinations in the bulk.

\section*{Results and Discussion}

\begin{figure}
  \centering
  \includegraphics[width=8.7 cm]{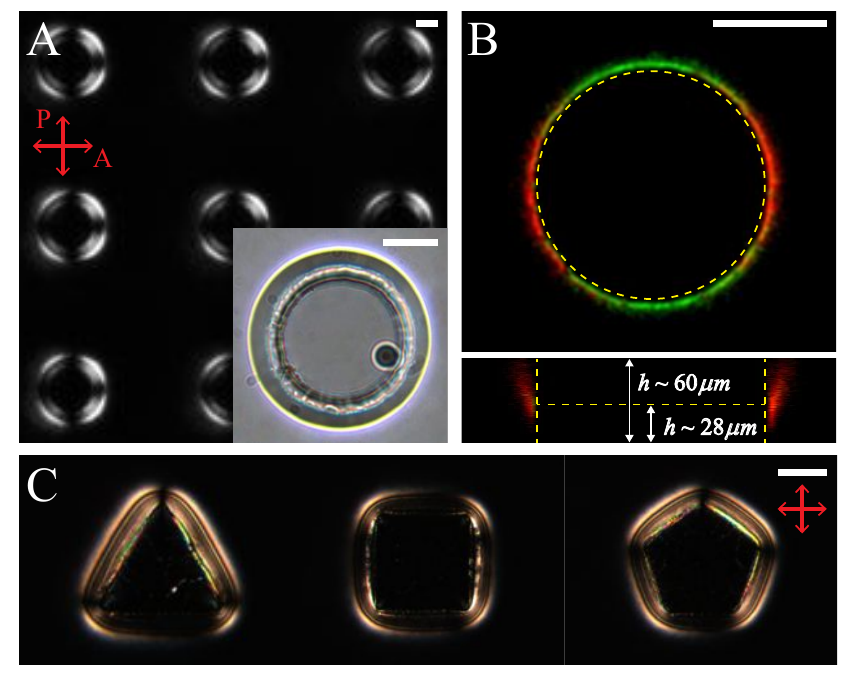}
  \caption{
Micropost induced bulk defect rings. (A) Polarized optical microscopy 			(POM) image of a micropost array where all surfaces have homeotropic 			anchoring resulting in defect rings that circumscribe each micropost. 
INSET: Bright field image of a single micropost where the bright line indicates the approximate lateral position of the defect loop. (B) Fluorescence confocal polarization microscopy (FCPM) image indicating the location of defects in an otherwise uniform director field. Top: A top view of the micropost. Bottom: A z-stack of FCPM images in which the maximum intensity represents the location of the defect core that occurs at approximately the post mid-height. (C) Disclination lines are dictated by the shape of the micropost as shown around triangular, square and pentagonal microposts. All scale bars are 50 $\mu$m.
 }
 \label{fig:fig1}
\end{figure}

\noindent We prepare a square array of cylindrical microposts and impose homeotropic anchoring at all surfaces, resulting in the appearance of a disclination ring around each micropost. The defect lines encircle the microposts and appear as bright lines in an otherwise black polarized optical micrograph shown in Fig.~\ref{fig:fig1}A. Under sample rotation, the dark regions remain black, confirming homeotropic anchoring, while bright field optical microscopy corroborates this observation (inset of Fig.~\ref{fig:fig1}A). When the sample is heated above the clearing point ($T_{NI}=34^\circ$C), the defect line is absorbed into the isotropic phase, resulting in a completely black image under crossed polarizers. Upon cooling back to the nematic phase (either slowly at 0.1 $^\circ$C/min or in contact with the room environment), the disclination reappears, confirming that the texture is due to the equilibrium elasticity of the NLC and the frustrated director field near the surface of the micropost. We determine the vertical position of the disclination line using fluorescence confocal polarizing microscopy (FCPM) and find that, for a wide range of post heights, the disclination occurs near the mid-height position (Fig.~\ref{fig:fig1}B). In addition, we deduce through FCPM that the disclination sits within $5\,\mu\hbox{m}$ of the micropost surface.  Though both bright field and polarized microscopy suggest a larger gap between the posts and defects, we believe this is an optical effect due to the curved interface and the birefringence of the LC.  Finally, by varying the cross section of the microposts we demonstrate that the disclination line bears the signature of the micropost shape as shown in Fig.~\ref{fig:fig1}C. 

\begin{figure}
  \centering
  \includegraphics[width=8.7 cm]{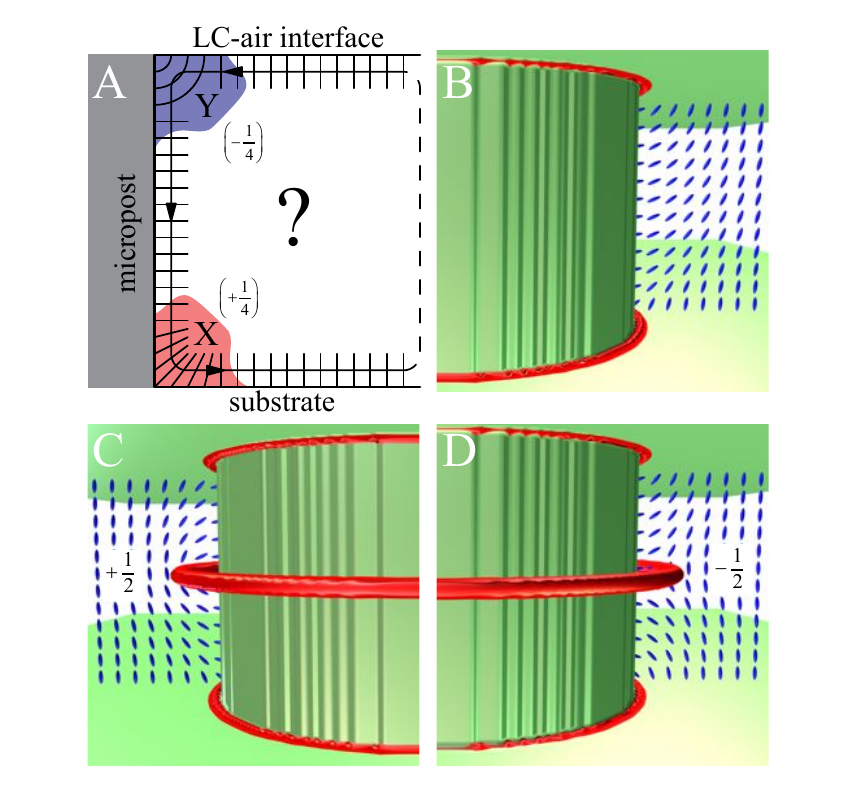}
  \caption{\label{fig:fig2}
Numerical and topological evaluation of the director field. (A) The system has two corners -- X and Y -- where the director field can choose between two winding senses.  (B-D) Director fields corresponding to the relative minima of the Landau-de Gennes free energy for a cylindrical micropost and planar interfaces, found numerically. Isosurfaces of $S=0.48$ are shown in red, where $S$ is the leading eigenvalue of $\mathbf{Q}$. Blue ellipsoids indicate the director field. The nematic is locally melted at the sharp corners.  (B) Opposite winding at the two corners precludes the possibility of a disclination in the bulk. (C) A bulk disclination with $+1/2$ (i.e., anticlockwise) winding number requires positive winding at both corners. (D) A bulk disclination with $-1/2$ (i.e., clockwise) winding number requires negative winding at both corners.  Since the numerics show that the (meta)stable states have azimuthal symmetry and that the director has no azimuthal component, we may think of these winding numbers as pseudo-charges.%
 }
\end{figure}

What does the observed disclination ring imply about the bulk director field and the boundary conditions associated with confinement?  To address this theoretically, we study a Q-tensor based, Landau-de Gennes (LdG) model of the nematic confined in micropost geometries. Using a finite difference scheme to minimize the LdG free energy \cite{Ravnik2009a} we determine the director field and defect locations. We use the unequal elastic constants measured for the elastic anisotropy of 5CB.  
We numerically model the micropost as a right cylinder bridging two planar interfaces and find four distinct local minima of the free energy: two in which there is no defect in the bulk (Fig. \ref{fig:fig2}B and its reflection through the horizontal), one with a disclination ring around the micropost with $+1/2$ winding geometry (Fig. \ref{fig:fig2}C), and one in which the disclination ring has instead a  $-1/2$ geometry (Fig. \ref{fig:fig2}D). All the minima exhibit axial symmetry and we find that the director field has no axial component as in Fig. \ref{fig:fig2}B-D.  It follows that we can discuss the topology of the texture in terms of {\sl two-dimensional} nematics corresponding to each radial slice, promoting the $\pm 1/2$ geometry into a pseudo-topological charge.  The disclination ring is topologically charged in the three-dimensional sense, as the director field profile is essentially constant along its contour \cite{janich1987topological, vcopar2011nematic, alexander2012colloquium}; we will henceforth focus exclusively on the two-dimensional pseudo-charge.

The total pseudo-charge can be calculated via the winding of the director around the boundary of the sample.  However, as our numerics demonstrate in Fig. \ref{fig:fig2}, the sharp corners at the cylinder-planar junctions require careful consideration of the topology in each slice.  Each corner can be resolved via a splay or bend texture leading to a director rotation of $\pm\pi/2$, respectively.  Thus the defect pseudo-charge in each radial slice is determined by the details of each junction.
When the winding sense is positive at corner X and negative at corner Y (Fig. \ref{fig:fig2}B) or {\sl vice versa}, there is no disclination ring in the bulk. On the other hand, a disclination ring of winding number $+1/2$ (Fig. \ref{fig:fig2}B) is observed at mid-height around the micropost when the two corners both have positive winding. Likewise, a disclination ring of winding number $-1/2$ (Fig. \ref{fig:fig2}C) is observed at mid-height around the micropost when the two corners both have negative winding. Thus, the experimental observation of a disclination in the bulk implies that the director winding has the same sign at both corners. The nematic degree of order is diminished locally at the corners due to the incompatible boundary conditions. It should be noted that the computed free energies of the states with the bulk disclination ring (Figs. \ref{fig:fig2}C and \ref{fig:fig2}D) are slightly higher than that of the states with no bulk defect (Fig. \ref{fig:fig2}B), whereas the bulk disclination is stable in experiment. A similar multistability due to sharp-cornered boundaries has been reported for nematics in square wells with planar anchoring \cite{tsakonas2007multistable, luo2012multistability}. There, the choice of winding sense at each corner gives rise to two optically distinct states, a useful situation for bistable displays. 

\begin{figure}
  \centering
  \includegraphics[width=8.7cm]{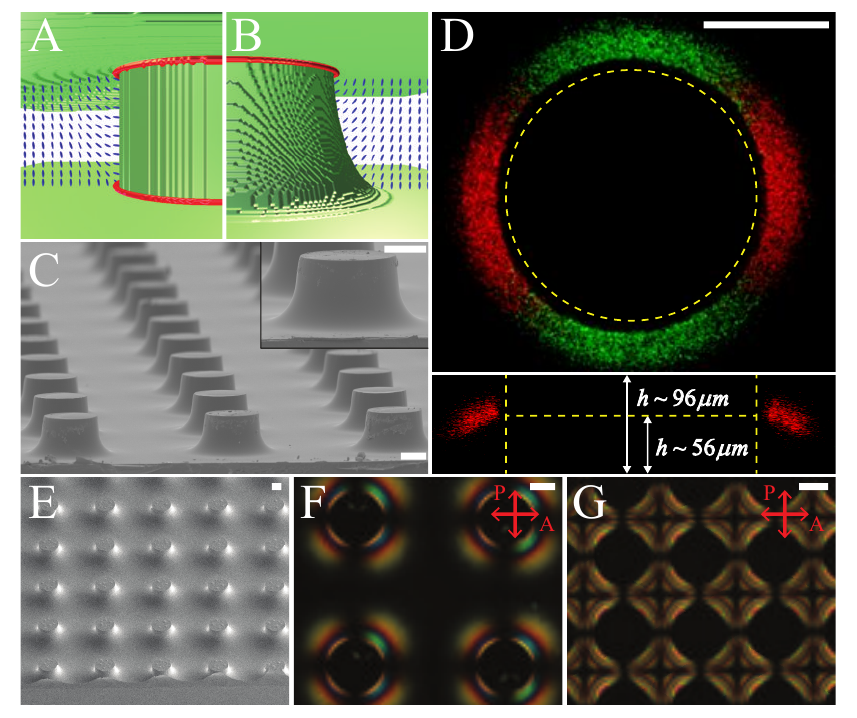}
  \caption{\label{fig:fig3}
The effect of surface curvature. (A) Where the 5CB-air interface curves upward to meet the micropost, LdG numerical modeling predicts that negative winding is favored. (B) Likewise, curvature at the bottom of the micropost favors positive winding; the surface defect in this case can be viewed as virtual, inside the micropost. (C) SEM image of microposts with a curved base and a corresponding (D) FCPM image detecting the presence of a disclination loop when the micropost array is filled with LC. Typically this defect will sit towards the upper half of the micropost. (E-G) Curved microposts tapered along their entire lengths, SEM image shown in (E), do not induce the formation of a bulk disclination ring (verified with FCPM), as is evident when viewed between crossed polarizers. The relaxation of the director to the vertical direction with increasing distance from the micropost is much more gradual in (F) than in Fig.~\ref{fig:fig1}A. Axial symmetry is lost at smaller micropost spacing (G). All scale bars are 50 $\mu$m.
 }
\end{figure}

In the numerical model, we can induce a local preference for a given winding sense by altering the geometry of either corner. For example, Fig. \ref{fig:fig3}A shows a 5CB-air interface curving upward to meet the micropost at a nonzero pinning angle as in the experiment. As a result, bend is favored over splay and the winding is negative. Conversely,  curvature at the bottom of the micropost favors positive winding, as shown in Fig. \ref{fig:fig3}B.  With this resolution of the sharp boundaries, the bulk disclination ring is not even metastable in the numerics.
To probe this experimentally, we replaced the sharp corner at X with a curved base to favor a positive winding sense ($+1/4$) as in Fig. \ref{fig:fig3}B; the surface pinning maintains the preference for negative winding at corner X.  In contrast with the numerics, we find that when the base of the micropost is slightly curved (Fig. \ref{fig:fig3}C) the bulk disclination persists but is pushed away from the base, closer to the 5CB-air interface. We verify the defect position using FCPM as shown in Fig.~\ref{fig:fig3}D and verify the stability of the defect by heating into the isotropic phase and re-cooling several times. Each time, upon cooling into the nematic phase, the defect re-forms. Further, the bulk disclination remains even when microposts are completely submerged in a thick nematic film. However, we also find that when the microposts are tapered all the way to the top (Fig.~\ref{fig:fig3}E-G), the bulk disclination ring fails to form (Fig.~\ref{fig:fig3}F and \ref{fig:fig3}G), as verified by FCPM, which is in agreement with the numerical results. Thus, the geometry of the boundary alone is not sufficient to predict the qualitative features of the bulk director field. We suspect that variations in surface chemistry, roughness, or relative magnitudes of anchoring strength are important for the stabilization of bulk defects. On the other hand, these findings demonstrate that sensitive control over the boundary shape offers the ability to tune the stable position of the disclination ring. { Lateral confinement of disclination rings by neighboring microposts holds promise as a way to further manipulate the defects, as shown by studies of nematic-filled capillaries \cite{de2007ringlike, de2007point}. However, simply decreasing the micropost center-to-center spacing (even down to $1.25$ times the micropost diameter) does not affect the existence or circular shape of the bulk disclinations.}

\begin{figure*}
  \centering
  \includegraphics[width=17 cm]{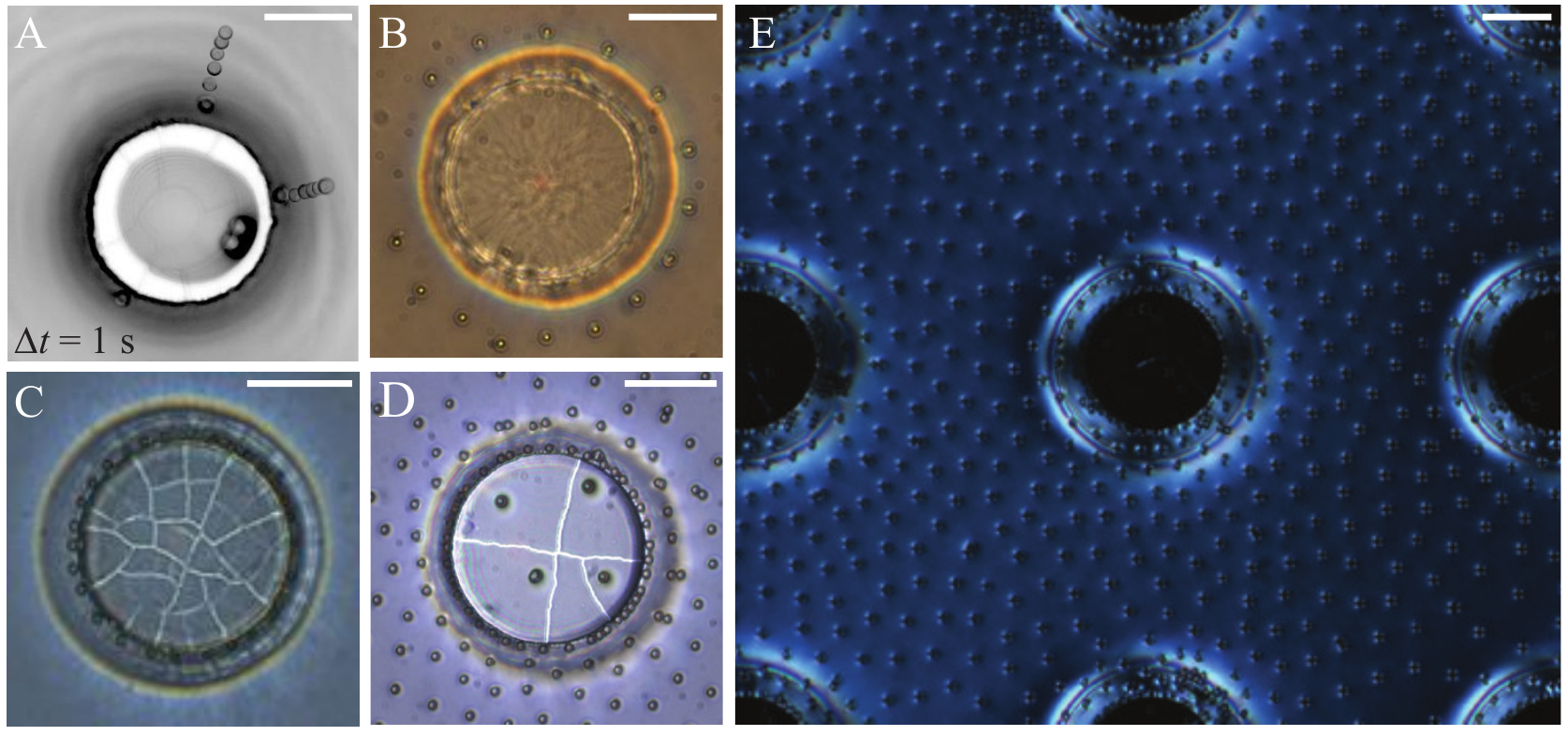}
  \caption{\label{fig:fig4}
Elastic migration of colloidal particles induced by bulk topological defects. (A) Time-lapsed images of the migration of spherical colloids towards a micropost ($ \Delta t$ = $ 1\,\mathrm{s} $). (B) A ring of colloids forms above a submerged micropost. (C) At moderate surface coverage, ordered rings assemble around the micropost due to attraction by the bulk defect and repel one another via long range interparticle repulsion. (D) As particle density continues to increase, radial assemblies evolve into hexagonal ordering until (E) highly ordered structures form at a very high surface coverage. All scale bars are 50 $\mu$m. 
 }
\end{figure*}

Armed with our careful characterization of the complexion, we now turn to our main interest: the interaction of the liquid crystal with colloidal particles.  Prior work has established that colloids are attracted to disclination lines, an effect that has been exploited in the bulk to form wire-like chains of colloidal particles along the defects \cite{Skarabot2008, Pires2007, Fleury2009, Bohdan2012}. Here however, we use the distortion field generated by the defect to {\sl remotely} steer particles trapped at the 5CB-air interface. When a $5\,\mu$m diameter silica microsphere with homeotropic anchoring is placed on the interface, it migrates radially toward the micropost until contact (Fig. \ref{fig:fig4}A). Though capillary interactions are known to induce particle migration along curved surfaces \cite{CavallaroPNAS}, capillarity effects are absent in this system since the interfacial distortions induced by the microspheres are negligible \cite{GharbiSoftMatter, Oettel2009}.  Further, when the system is heated above the nematic-isotropic transition to annihilate the disclination lines we no longer observe migration and thus conclude that the observed migration must be orchestrated solely by the elastic director field. In other words, the elastic distortion created by the spherical particle is interacting with the bulk nematic texture.  As noted in \cite{GharbiSoftMatter}, however, the distortions made by the particles are quadrupolar in nature and die off rapidly away from the surface.  Thus we expect that the colloids will only interact with large director deformations. As shown in Fig.~\ref{fig:fig2}B-D, we can imagine three potential sources of large director distortions which include a region of splay gradient along the interface due to its meniscus; a defect line near the interface at the upper corner Y (as in Fig.~\ref{fig:fig2}A); and the defect ring in the bulk. We can eliminate the first potential source by submerging the micropost in the liquid crystal so that it no longer intersects the nematic-air interface. Yet, even without distortion of the interface, migration occurs. When several particles are present at the interface, they form a ring above the immersed micropost mimicking the bulk disclination (Fig.~\ref{fig:fig4}B).  The second potential source is shown to be negligible by changes to the micropost geometry discussed below, which alter the speed of colloidal migration without changing the defect at corner Y. These observations confirm that the assembly of particles at the free interface is, in fact, driven by the third source, the defect ring within the bulk liquid crystal.

As we increase the surface density of the microspheres, a single ring 
of colloids around each micropost (Fig. \ref{fig:fig4}B and \ref{fig:fig4}C) gives way to complex structures that form owing to the balance of attraction to the micropost and the long range interparticle elastic repulsion. This competition results in highly ordered structures as shown in Fig. \ref{fig:fig4}D and Fig. \ref{fig:fig4}E that nucleate from the post and evolve with distance away from the post into a hexagonal lattice.

\begin{figure}
  \centering
  \includegraphics[width=8.7 cm]{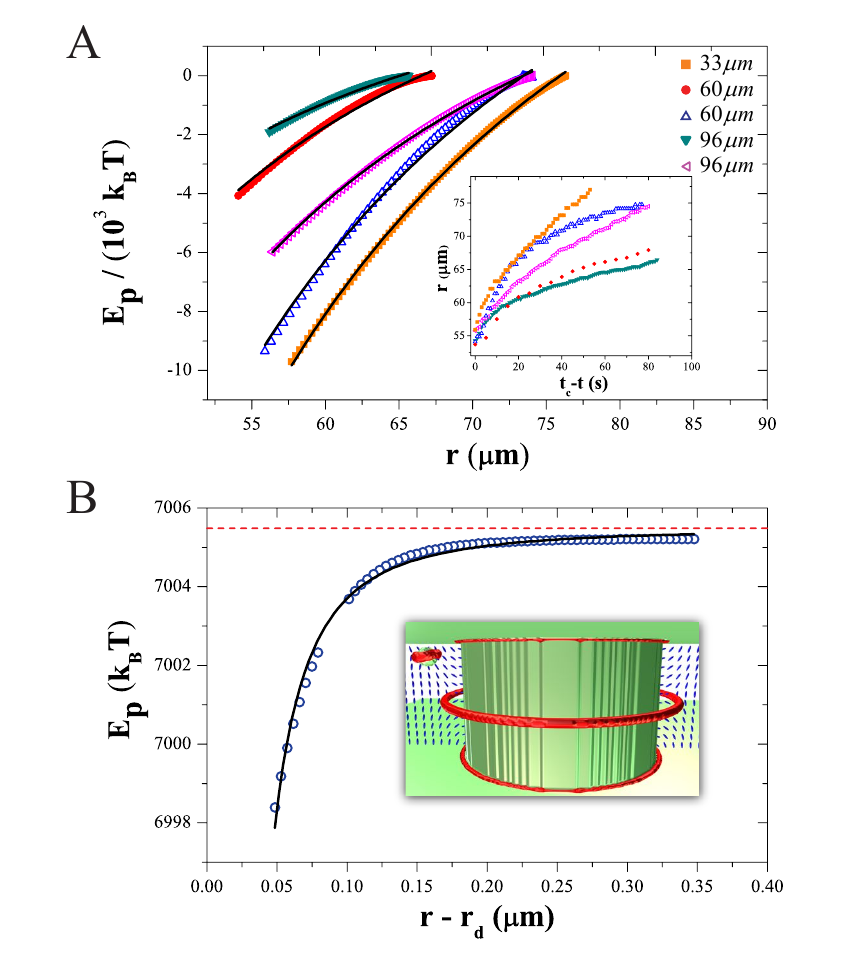}
  \caption{\label{fig:fig5}
Elastic potential. (A) Migration of colloids implies bulk-colloid interaction energies on the order of $1\times10^{4}$ $kT$ and follow $ E_p \propto r^{-2} $ (solid curves), where $E_p$ is the inferred elastic potential and $r$ is the radial distance in the horizontal plane from the center of the micropost. As the height of a micropost is increased (different curves), attractions become weaker due to an increased separation between particles and the bulk defect. INSET: Migration rates are faster for microposts with base curvature compared to those of comparable height that have a sharp corner at the base (corner Y in Fig.~\ref{fig:fig2}A). This is due to the tendency of the bulk disclination to position itself closer to the free interface for microposts with curved bases. Closed symbols: microposts with curved base. (B) Numerically modeled energy of a colloidal sphere approaching a micropost capturing the $ E_p \propto (r-r_d)^{-2} $ dependence of the quadrupolar colloid interacting with a disclination ring. The red dashed line represents the asymptotic value for $E_{p}$ at large distances and $r_d$ is the radius of the micropost's disclination ring. INSET: Representative image of the numerical modeling.
 }
\end{figure}

Since the particles migrate in creeping flow, we can estimate the interaction energy by balancing the elastic and viscous forces. We track particle trajectories and use drag coefficients implied by  the Brownian fluctuation of particles at the interface and the Stokes-Einstein relation.  We infer elastic interaction energies $E_p\approx 10^{4}$ $kT$ (Fig. \ref{fig:fig5}A) and find that the energy falls off as the inverse square of the radial position from the post center.   Additionally, we test the elastic potential numerically by moving a small spherical colloid with a quadrupolar defect at fixed height near the top surface toward a micropost with a defect ring. As shown in Fig. \ref{fig:fig5}B, the elastic energy is well described by $E_p = E_0 - \alpha (r-r_d)^{-2}$, where $r_d$ is the radius of the disclination ring around the micropost and $E_0$ and $\alpha$ are fitted coefficients. The dependence of this potential on distance from the disclination, rather than from the center or edge of the micropost, for example, further underscores the role of the disclination as the principal cause of the elastic attraction. 

Finally, we find that as the height increases the rate of migration decreases (inset of Fig. \ref{fig:fig5}A). Since the disclination line is always at or below the micropost mid-height for un-tapered microposts, the reduced strength of interaction follows from the increased separation between the microspheres at the surface and the defect ring in the bulk. We confirm the latter result by measuring the variation of the elastic force magnitude $f_p$ for micropost heights of 33 $\mu$m, 60 $\mu$m, and 96 $\mu$m at fixed radial position of the particle $r=62$ $\mu$m. We find that $f_p$ decreases with values of 1.39 pN, 0.633 pN, and 0.406 pN, respectively. Similarly, when the micropost base geometry is changed from a ``sharp'' base to a curved base, the defect rings are closer to the interface, and the migration rates increase.  Together these observations offer additional confirmation of the direct interaction of the bulk disclination and the colloids.

To summarize, we have demonstrated that bulk defects can be generated in confined liquid crystals and guide the assembly of remote colloidal particles into ordered structures through nematic elasticity.  The colloids assemble to mimic the defect structure in the bulk, in this case, a ring around the micropost.  Even when the microposts are submerged under liquid crystal, the elastic effect of the trapped topological defects remotely organizes the colloids on the surface, allowing us to probe the interior through ``colloid-aided tomography.''   At high surface coverage, attraction to the defect ring is coupled with long range interparticle repulsion and leads to highly ordered structures that nucleate radially outward from microposts. Coupled with the noted ability to manipulate the nematic director through flow, fields, and functionalization, our work paves the way to dynamically tunable assemblies at fluid interfaces.



{ \fontsize{10pt}{13pt}\selectfont 
\section*{Materials and Methods}

\noindent \textbf{Liquid Crystal 5CB.} We use 4-cyano-4'-pentylbiphenyl (5CB, Kingston Chemicals Limited), a thermotropic liquid crystal with a nematic phase between $ 18^{\circ} C \le T \le 34^{\circ} C $. \\
\\
\noindent \textbf{Supporting 5CB-Air Interfaces in Micropost Arrays with Homeotropic Anchoring.} KMPR negative photoresist (Microchem Corp.) is used to fabricate micropost arrays on transparent quartz substrates. The cylindrical microposts have nominal radii $R=50\,\mu\hbox{m}$, heights ranging from  $33\,\mu\hbox{m}\le h\le 96\,\mu\hbox{m}$, and are positioned in square arrays with a pitch $p=400\,\mu\hbox{m}$. The micropost height is tuned by changing spin speed when depositing the photoresist while different micropost cross-sections are created by varying the photomask design. Microposts with curved bases are achieved by spin coating SU-8 2002 on a micropost array at speeds between 1500 and 6000 rpm and then developing the epoxy using standard SU-8 lithographic procedures. All substrate surfaces are treated to induce strong homeotropic anchoring by sputtering a  $30\,\hbox{nm}$ film of chromium using a home built sputtering system and then incubating the samples in a 3\% weight solution of 3-(Trimethoxysilyl)propyl-octadecyldimethylammonium chloride, (DMOAP, Sigma Aldrich) in a 9:1 mixture of ethanol:water. Samples are rinsed and heated at 110$^\circ$C for 1 hour to complete the substrate treatment \cite{Bullet}. Untreated substrate surfaces have degenerate planar anchoring. To confine the LC film, the micropost array substrate is heated to 60$^\circ$C and 5CB  is applied by spin-coating. At the top edge of each micropost, the nematic-air interface is pinned with a typical angle $15^\circ\le\psi\le20^\circ$ measured with respect to the horizontal using a goniometer. \\
\\
\noindent \textbf{Preparing and Depositing Microspheres.} Dry silica microspheres of nominal diameter $5\,\mu$m (Polysciences, Inc.) are treated to induce strong homeotropic anchoring using the same chemistry described above for the microposts. Particles are then dried in a vacuum oven overnight at 110$^\circ$C and subsequently aerosolized in a closed container with compressed air. Once aerosolized, the substrate supporting the 5CB-air interface is introduced and particles are effectively deposited with minimal aggregation. \\
\\
\noindent \textbf{Optical Characterization.} We characterize our system using an upright microscope in transmission mode equipped with crossed polarizers (Zeiss AxioImager M1m) and a high resolution color camera (Zeiss AxioCam HRc). Fluorescence confocal polarization microscopy (FCPM) images are obtained using an inverted IX81 microscope equipped with an FV300 confocal scanbox and a half wave plate between the objective and filter cubes to rotate the polarization of the scanning laser. FCPM allows us to determine director orientation of LC molecules doped with a 0.01\% wt. dye N,N'-Bis(2,5-di-tert-butylphenyl)-3,4,9,10-preylenedicarboximide (BTBP, Sigma Aldrich). At these low concentrations, dye molecules align along the axis of the LC molecules while preserving the properties of 5CB and fluoresce when aligned parallel to the excitation light polarization vector.  \\
\\
\noindent \textbf{Numerical modeling.} Our numerical modeling follows the Landau-de Gennes numerical approach reviewed in \cite{Ravnik2009a}. Our finite difference scheme employs a regular cubic mesh on which is defined a traceless, symmetric, rank-2 tensor $\mathbf{Q}$, from which the nematic director is recovered as the eigenvector corresponding to the leading eigenvalue $S$. The LdG free energy density is a sum of two terms,
\begin{align*}
f_{\mathrm{phase}} &= \tfrac{1}{2} A Q_{ij} Q_{ji} + \tfrac{1}{3} B Q_{ij} Q_{jk} Q_{ki} + \tfrac{1}{4} C \left(Q_{ij} Q_{ji}\right)^2, \\
f_{\mathrm{grad}} &=\frac{1}{2} L_1 \frac{\partial Q_{ij}}{\partial x_k} \frac{\partial Q_{ij}}{\partial x_k} + \frac{1}{2} L_2 \frac{\partial Q_{ij}}{ \partial x_j} \frac{\partial Q_{ik}}{\partial x_k}  \\ &+ \frac{1}{2} L_3 Q_{ij} \frac{\partial Q_{kl}}{\partial x_i} \frac{\partial Q_{kl}}{\partial x_j}
\end{align*}
with summation over repeated indices. In $f_{\mathrm{grad}}$, we use $L_1=3.8\times10^{-12}$ N, $L_2=5.3\times10^{-12}$ N, and $L_3=5.3\times10^{-12}$ N to model 5CB with elastic constants $K_1=0.64\times10^{-11} $ N, $K_2=0.3\times10^{-11}$ N, $K_3=1\times10^{-11}$ N \cite{kleman2002soft}. It is necessary to include a third-order term in $f_{\mathrm{grad}}$ in order to have $K_1\neq K_3$ \cite{Ravnik2009a}. From Ref. \cite{Ravnik2009a}, we take typical values for the material constants of 5CB: $A=-0.172\times10^6\;\mathrm{J/m^3}$, $B=-2.12\times10^6\;\mathrm{J/m^3}$, $C=1.73\times10^6\;\mathrm{J/m^3}$. From a random or specified initial configuration, we minimize the LdG free energy over $\mathbf{Q}(\vec{x})$ using a conjugate gradient algorithm from the ALGLIB package \cite{ALGLIB}. The mesh spacing corresponds to 4.4 nm. The microposts modeled have diameter 440 nm and height 264 nm. The simulation box has length 1320 nm in both horizontal dimensions.  

We model the anchoring strength on all the substrate surfaces as infinite by imposing a fixed, uniaxial $\mathbf{Q}$ at the boundaries. The infinite anchoring strength approximation is reasonable because 5CB near DMOAP treated surfaces experiences an anchoring strength $W\approx 10^{-2}$ N/m \cite{Skarabot2008}, so the anchoring extrapolation length $\xi_W=K/W$ is a few nanometers, close to the nematic coherence length $\xi_N\sim\sqrt{L_1/A}$. We have verified that finite anchoring strength in the strong anchoring regime does not significantly change the results using a Rapini-Papoular-type surface potential as used in Ref. \cite{Ravnik2009a}. Periodic boundary conditions in the horizontal directions are used to simulate a square array of microposts.\\
\\
}


\begin{acknowledgments}
The authors acknowledge support from National Science Foundation (NSF) through MRSEC Grant DMR11-20901. This work was supported in part by a gift from The Mark Howard Shapiro and Anita Rae Shapiro Charitable Fund and by NSF Grant PHY11-25915 under the 2012 Kavli Institute for Theoretical Physics miniprogram ``Knotted
Fields''. DAB was supported by the NSF through a Graduate Research Fellowship under Grant No. DGE-1321851.  This work was partially supported by a Simons Investigator award from the Simons Foundation to RDK. DAB and RDK thank the Isaac Newton Institute for their hospitality while this work was completed.
\end{acknowledgments}







\newpage 

\end{document}